# A Connected Enterprise- Transformation Through Mobility And Social Networks

Jitendra Maan

Tata Consultancy Services, TCS Towers, 249 D & E, Udyog Vihar Phase IV
Gurgaon, Haryana, India – 122001
`jitendra.maan@tcs.com`

## ABSTRACT

*Due to rapid changes in business dynamics, there is a growing demand to encourage social conversations/exchanges and the ability to connect and communicate with peers, partners, customers and other stakeholders anytime, anywhere which drives the need of mobile-enable, the existing enterprise applications.*

*This paper highlights a distinct set of needs and key customer challenges that must be considered and addressed for deployment of Social Collaboration applications and Mobility services in enterprises. It not only addresses the Critical Success Factors for enterprise mobility enablement but also outlines the unique business requirements to rapidly create social collaboration culture and the discipline of turning social data into meaningful insights to drive business decisions in real-time.*

*Moreover, the paper emphasizes on developing composite offerings on social enterprise and Mobile networks that not only offer the value proposition in terms of financially oriented results, but also help customer to maximize return on investment (ROI).*

## KEYWORDS

*Enterprise Mobility, Mobile Solutions, Social Networks, Social Enterprise, Social Collaboration*

## 1. INTRODUCTION

Most enterprises today face a common set of challenges when it comes to integrating their mobile workers into their enterprise-wide collaboration networks. Both, Mobility and social collaboration networks are transforming the industry by finding their feet in almost every industry sector and they have a profound impact on both customers as well as employees.

An interesting trend being observed currently is that social networking and mobility solutions come together to the point where social collaboration networks fundamentally focus on building online communities of people and Mobility becomes the most preferred delivery channel not just to improve the return on investment, but also to expand global reach and improve operational efficiency of the enterprise worker.

However, it has become imperative to make a clear distinction between "mobile enablement" and "mobile social network-enablement" – whereas the former simply means the rendering of an existing web application on a mobile channel, whereas mobile social network-enablement means embedding on-demand social networking and collaboration capabilities into the application to increase its utility. Today, there are several business use cases that either leverage Mobility platform to host Social networks or they would create scalable Mobile applications available Social networking platforms.

## 2. PENETRATION OF SOCIAL COLLABORATION AND MOBILITY-ENABLEMENT ON RISE

Fundamentally, social networking and collaboration solutions are growing across business portfolios, driven by the need to increase productivity and improve decision making, with increased near real-time access through Mobile channels and such adoption helps to improve internal employee interaction, customer collaboration, network building and information sharing.

Realizing the importance of enterprise mobility as a strategic priority, enterprises should certainly empower their business processes with the enterprise-wide deployment of mobile applications. Enterprise Mobility has the potential to fundamentally transform enterprises, their business value chains and markets. Organization will have a clear vision to explore how to operate in ubiquitous computing eco-system when location constraints are obliterated and mobility becomes the key endpoint for delivery.

Organizations should not consider social network and collaboration as a stand-alone activity. Most of the customers who have laid out social media solutions, have seen the maximum benefits by integrating social media with customer experience management and have developed the processes to access it on Mobile devices appropriately. Mobile services and social collaboration applications not only provide a rich, collaborative, social experiences to users, but helps to foster collective intelligence—the "wisdom of the crowds"—and evolving the new way to get insights, opinions, perceptions and also open up conversation with communities.

## 3. KEY BARRIERS TO ORGANIZATION SUCCESS

Fundamentally, social networking and collaboration solutions are growing across business portfolios, driven by the need to increase productivity and improve decision making, with increased near real-time access through Mobile channels and such adoption helps to improve internal employee interaction, customer collaboration, network building and information sharing.

### 3.1. CIO Perspective – Roadblocks to IT growth

Social computing and networking has caused a dramatic evolution in the way people collaborate and interact via the Internet. Essentially, social computing represents the collection of technologies that gather, process, compute, and visualize social information. This new social structure has emerged creating an array of loosely integrated social components.

A competitive and agile business environment continue to force CTOs/CIOs to optimize their IT investments and address the challenge of achieving more with less capital, by developing a clear understanding on what to socialize, when to socialize, as well as how to develop and execute mobility strategies in the context of their overall business eco-system.

### 3.2. Key Challenges – Enterprise Mobility Adoption

Mobile paradigm is different from the normal client-server based application development. The bottom line is that mobile application development is new to most organizations and comes with unique challenges.

The key customer challenges/pain areas include the following:

- Real Time access to critical information is not available when needed
- Increasing customer expectation of immediate response to problem resolution
- Lack of adoption of globally available mobility solutions
- Re-use/utilize existing web-oriented infrastructure
- Scalable and flexible Mobility solutions with changing business needs
- Lack of standardization in Mobile offerings from Mobile platform vendors.
- Fragmented mobile platforms and diversity issues. Eg Multiplicity of

International Journal of Managing Information Technology (IJMIT) Vol.4, No.3, August 2012

- Connected devices,
- Operating Systems,
- Different flavor of Mobile Browsers,
- Different Form factors and input methods

### 3.3. Key Challenges – Social networking and collaboration Platform Enablement

Enterprise-wide collaboration is not only a business objective, but it has become imperative for enterprises to remain competitive and provide actionable information in a timely manner. The typical challenges faced by organizations in enablement of Social Collaboration as below -

- **Lack of transparency** – Lack of access to the right information, to the right people, at the right time
- **Slow and inefficient decision making** - Business leaders frequently make critical decisions without the information they need
- **Customer churn**–Lack of analysis on customer needs and grievances which will lead to customer dissatisfaction
- **Costly and ineffective customer service** – Organizations are able to operate only in reactive mode to customer problems as and when the service/support requests are logged
- **Ineffective and costly sales and marketing** – Lack of analysis on customer buying behaviors, patterns and grass root needs
- **Slow customer acquisition and growth** –Lack of data mining capability, for generating new customer leads or opportunities
- **Locating resources become a nightmare** –Lack of ability to find the right resource across all data sources and content types and no consistent view across all business applications and data sources

## 4. BUSINESS IMPACT ACROSS LINE-OF-BUSINESS APPLICATIONS

### 4.1. Mobility Impact on Business-critical Applications

Most companies, regardless of their size and location, today face a common set of challenges when it comes to integrating their mobile workers into the enterprise's processes. They have invested a lot in implementing ERP and CRM systems. Most companies will adopt Mobile technologies gradually, without ripping and replacing their core enterprise resource planning (ERP) Platforms.

However, these applications are rendered useless as soon as a mobile worker steps out of the office, because much of the activity of customer-facing or mobile professionals is dependent on timely and accurate access to the enterprise information and processes. For example, CRM systems have been implemented to support the activities of sales and services people, but because these people are mostly mobile, the value of the CRM implementations is dramatically reduced without the necessary mobility support. The figure below shows the level of Mobility impact across business Applications.

International Journal of Managing Information Technology (IJMIT) Vol.4, No.3, August 2012

Figure 1.  Impact of Mobility across Application Portfolio

## 4.2. Business Impact of Social Networking and Collaboration

Some of the key drivers for Social Networking that impact the bottom line across different line of businesses, are explained below –

- **Collaboration by building Communities -** An interactive collaboration and engagement eco-system to facilitate sharing knowledge around products, services, technologies and business issues by tapping unstructured data
- **Customer care and Insights -** Filtering the data flow, prioritize customer conversations, analyze social interaction patterns, behaviors and accelerate decisions based on those unheard insights.
- **Product and Service Innovation–** To create a continuous stream of products, solutions and services leveraging an ecosystem of best-of-breed ideas, technologies and capabilities from Open Innovation partners.
- **Knowledge Management –** Create knowledge repositories by crowd sourcing rich pool of innovative ideas/concepts, views, opinions, customer feedback, expert comments leveraging an interactive community ecosystem of Customers, open innovation partners and employees.
- **Improve Self Service –** Customer self-service business-driven solution.
- **Expertise Management -** Finding and connecting people, teams and expertise, collaborating socially

The figure below examines various Social Networking driving behaviours that have profound impact on enterprises.

92

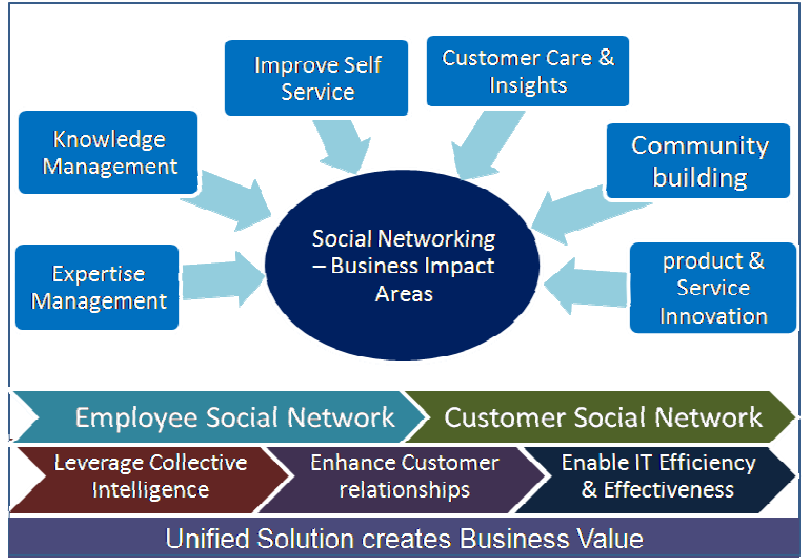

Figure 2. Business Impact of Social Networks

## 5. COMPOSITE OFFERING ON SOCIAL NETWORKING AND MOBILITY ENABLEMENT ACROSS ENTERPRISES

With the proven industry experience, Tata Consultancy Services (TCS), as an IT Services Vendor, bring to the customers, an integrated composite offering which brings together various solution enablers to harness the power of both social media and cross platform Mobility augmented by right set of tools to optimize the customer IT investment. Such solution offerings not only push the boundaries of its solution portfolio, but utilizing the flexibility of social networking platforms to integrate with other systems and enterprise applications within the organization. Such integrated composite offering solutions help organizations to reach beyond existing customers and will become a source of collaboration on an on-going basis with their stakeholders including employees, partners and customers. Most of the enterprise customers are applying the concepts and lessons learned from social networking and collaboration to connect people and enhance business interaction.

Besides the benefits of mobility enablement, a full-fledged platform, sometimes may not be addressing the need for every organization. In such a scenario, opting for enterprise mobility services from a hosted/managed services vendor is a viable choice. Now that several service providers are moving to the cloud for better service penetration and cost optimization, this option should certainly be explored by enterprises to improve overall customer satisfaction and set themselves apart from their competition.

## 6. RECOMMENDATIONS AND KEY PERFORMANCE INDICATORS

### 6.1. Enterprise Mobility – Value Proposition

Organizations need a strong mobility platform to gain the benefits from enterprise mobility. Enterprise Mobility will not only open up new channels for collaboration but provides the means for accessing the critical customer data on the go - Anytime, anywhere. Based on my study with customers from different industry clusters, I came up with the fact that a sharp distinction has to be drawn between creating a technology-focused mobile strategy, which refers to helping employees find the right devices, platform, and applications, and a business-focused strategy for mobility, which analyzes how mobility will affect various stakeholders. Based on the informal study with several enterprises, the following conclusions are drawn on the possible impact of enterprise mobility and value proposition to potential organizations in terms of tangible and financially oriented results (Refer the figure '3' below).

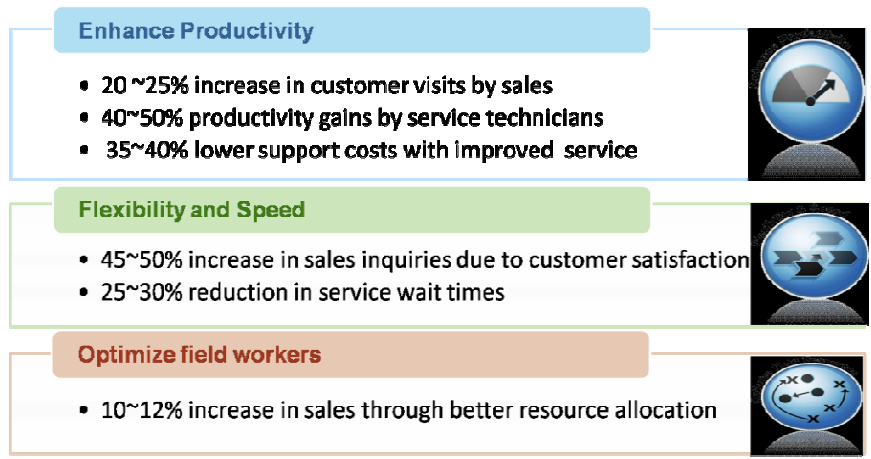

Figure 3. Enterprise Mobility – Tangible and Financial Results

The Key recommendations, that may be considered while developing the Mobile applications are given below –

- Understand the customer mobile platform preferences
- Select the platforms that are used by key target segments
- Have a clear business case for multi-platform mobile development
- Keep as much of the logic in the network-side as possible
- Mobility solutions must utilize the existing Web-oriented infrastructure by reusing Web Services architecture as much as possible.
- Test the mobile application on actual devices, with real customers, using it in the typical and expected real-world contexts/situations.
- Develop rich enterprise mobile applications while minimizing costs using -
    - Right architectural approach
    - Deep knowledge of mobile as a medium
    - Competitive delivery models

## 6.2. Social Collaboration and Networking – Key Result Areas

Following are the key result areas in terms of absolute business gains achieved with Enterprise Social Collaboration solutions implemented across organizations –

- 25~30% increase in speed of access to knowledge
- 20~25% faster access to Expertise
- Improved business decisions –
    - 12~15% increase in successful innovations and ideas around products and Services
- Increased Customer satisfaction and loyalty
    - 10~15 % higher customer satisfaction through Self-service Social Collaboration
    - 8~10% higher customer loyalty
- Improved sales process
    - 5~8% increase in revenue
    - Sales Lead generation improved by 8~12%

Besides this, there are a number of key success factors that organizations have to consider and plan accordingly to maximize their return on investment and business benefits to achieve their goals –

- **Define a strategic enterprise collaboration initiative** - An overarching enterprise collaboration plan needs to be initiated in support of enterprise-wise strategy and vision.
- **Building success measures and metrics into Social Collaboration initiative** –Return on Investment should be clearly measurable and should align metrics with business goals
- **Integrate Core Social Collaboration capabilities with enterprise resources** – In order to deliver an integrated business capability, enterprise information like user profiles, communities should be centrally integrated with identity management and corporate policy systems
- **Social Collaboration integrated into mission critical enterprise applications and processes** - Bottom-line business value is realized while Social Collaboration initiatives are integrated with enterprise applications and processes.

## 7. CONCLUSIONS

It is now clear that convergence of social networks and mobility deployment would not only help to improve the efficiency of on-the-move workers, but also result in creating a dynamic ecosystem of online services, environments and applications. Social networks are becoming unique touch points to engage communities, initiate conversations and developing innovative ideas. Organizations has started leveraging social media to build a knowledge ecosystem with customers, prospects, employees and also making better use of feedback and coordinate their actions and exchange information anytime, anywhere using any medium.

Companies that can create or participate in a collaborative network and organize themselves to best leverage the benefits will enjoy a competitive advantage. Companies of all sizes can start by setting up a network of internal experts, suppliers, partners and customers. There are even technology platforms that can be employed to facilitate this collaboration so that you can get started immediately if you want.

Social business collaboration allows companies to exchange thoughts and ideas in a way that is integrated with business processes, and promises to become a very strong business tool. The greatest benefit will come from combining the collaborative power of social networks with the Mobility Strategy.

As organizations begin to see success stories and case studies take shape, they can begin to plan social computing investments that involve customers, partners, and external communities. Organizations can also seek to leverage the relationship between business decision makers and IT to adopt or develop richer sets of social and Mobile tools that help to enable social computing both within and outside of the firewall.

## REFERENCES

[1] Rahul C. Basole, Enterprise mobility: Researching a new paradigm, Information Knowledge Systems Management 7 (2008) 1–7, IOS Press.

[2] S.J. Barnes, Enterprise Mobility: Concept and Examples, International Journal of Mobile Communications 1(4) (2003), 341–359.

[3] John Saeteras, Mobile Web vs. Native Apps, Revisited, 2010

[4] "Mobilizing Applications for the Enterprise," Mobile Enterprise Magazine, June 2011: http://tiny.cc/mznfc

[5] Dana Moore, Raymond Budd and Edward Benson, "Professional Rich Internet Applications: AJAX and Beyond", Wrox Press © 2007 Citation

[6] Designing for the Mobile Web, Sitepoint -http://www.sitepoint.com/designing-for-mobile-web

[7] Stuart J. Barnes, International Journal of Mobile Communications, Inderscience Publishers, Volume 1 Issue 4, October 2003

[8] Fang Wang and Yaoru Sun, (August 2008) Self-organizing peer-to-peer Social Networks. Computational Intelligence, Vol. 24, No. 3.

[9] Jennifer Golbeck, Matthew Rothstein, (2008) "Linking Social Networks on the Web with FOAF: A Semantic Web Case Study. In Proceedings of the Twenty-Third Conference on Artificial Intelligence (AAAI'08)", Chicago,Illinois, July 13–17, 2008m

## Author

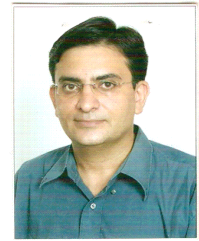

Jitendra Maan, a versatile IT Professional having a total of more than 15 years of IT experience and currently working with Tata Consultancy Services Limited in a leading role to drive Mobility & Social Computing Solutions at HiTech Solutions Central (HTSC) Group. Jitendra practices technology consulting, enterprise architecture and evangelizes Mobility-driven cloud solutions initiatives within TCS and has successfully delivered technology solutions for globally distributed clientele. Jitendra is certified in Project Management (CIPM) by the Project Management Associates (PMA), India and has successfully achieved the standards of TOGAF 8 Certification program. Jitendra has a proven track record of sharing technology thought leadership in various international conferences and also presented his research work on international platforms.